\documentclass[aps,prl,twocolumn,showpacs,reprint,superscriptaddress,nofootinbib]{revtex4-1}

\pdfoutput=1

\usepackage{graphicx}
\usepackage{xspace}

\RequirePackage{lineno}
\begin{document}
\title{
First Muon-Neutrino Disappearance Study with an Off-Axis Beam 
}

\newcommand{\superk}           {Super-Kamiokande\xspace}       
\newcommand{\nue}                {$\nu_{e}$\xspace}
\newcommand{\nuebar}           {$\bar{\nu}_{e}$\xspace}
\newcommand{\numubar}           {$\bar{\nu}_{\mu}$\xspace}
\newcommand{\numu}             {$\nu_{\mu}$\xspace}
\newcommand{\nutau}             {$\nu_{\tau}$\xspace}
\newcommand{\nux}                {$\nu_{x}$\xspace}
\newcommand{\numunue}       {$\nu_{\mu} \rightarrow \nu_{e}$\xspace}
\newcommand{\numunux}       {$\nu_{\mu} \rightarrow \nu_{x}$\xspace}
\newcommand{\numunutau}    {$\nu_{\mu} \rightarrow \nu_{\tau}$\xspace}
\newcommand{\tonethree}       {$\theta_{13}$\xspace}
\newcommand{\tonetwo}         {$\theta_{12}$\xspace}
\newcommand{\ttwothree}       {$\theta_{23}$\xspace}
\newcommand{\ssttmue}          {$\sin^2 2 \theta_{{\mu}e}$\xspace}
\newcommand{\sstonethree}    {$\sin^2 2 \theta_{13}$\xspace}
\newcommand{\ssttwothree}    {$\sin^2 2 \theta_{23}$\xspace}
\newcommand{\msqonetwo}   {$\Delta m^2_{12}$\xspace}
\newcommand{\msqonethree} {$\Delta m^2_{13}$\xspace}
\newcommand{\msqtwothree} {$\Delta m^2_{23}$\xspace}
\newcommand{\absmsqtwothree} {$|\Delta m^2_{23}|$\xspace}
\newcommand{\msqmue}        {$\Delta m^2_{{\mu}e}$\xspace}
\newcommand{\msqmumu}     {$\Delta m^2_{\mu\mu}$\xspace}
\newcommand{\enu}               {$E_{\nu}$\xspace}
\newcommand{\pmu}              {$p_{\mu}$\xspace}
\newcommand{\amome}         {$E_{e}$\xspace}
\newcommand{\evis}              {$E_{vis}$\xspace}
\newcommand{\pizero}           {$\pi^{0}$\xspace}
\newcommand{\pizerogg}       {$\pi^{0}\to\gamma\gamma$\xspace}
\newcommand{\degree}      {$^\circ$\xspace}

\newcommand{\INSTC}{\affiliation{University of Alberta, Centre for Particle Physics, Department of Physics, Edmonton, Alberta, Canada}}
\newcommand{\INSTDF}{\affiliation{National Center for Nuclear Research, Warsaw, Poland}}
\newcommand{\INSTEE}{\affiliation{University of Bern, Albert Einstein Center for Fundamental Physics, Laboratory for High Energy Physics (LHEP), Bern, Switzerland}}
\newcommand{\INSTFE}{\affiliation{Boston University, Department of Physics, Boston, Massachusetts, U.S.A.}}
\newcommand{\INSTD}{\affiliation{University of British Columbia, Department of Physics and Astronomy, Vancouver, British Columbia, Canada}}
\newcommand{\INSTBR}{\affiliation{Brookhaven National Laboratory, Physics Department, Upton, New York, U.S.A.}}
\newcommand{\INSTGA}{\affiliation{University of California, Irvine, Department of Physics and Astronomy, Irvine, California, U.S.A.}}
\newcommand{\INSTI}{\affiliation{IRFU, CEA Saclay, Gif-sur-Yvette, France}}
\newcommand{\INSTCI}{\affiliation{Chonnam National University, Institute for Universe \& Elementary Particles, Gwangju, Korea}}
\newcommand{\INSTGB}{\affiliation{University of Colorado at Boulder, Department of Physics, Boulder, Colorado, U.S.A.}}
\newcommand{\INSTFG}{\affiliation{Colorado State University, Department of Physics, Fort Collins, Colorado, U.S.A.}}
\newcommand{\INSTCJ}{\affiliation{Dongshin University, Department of Physics, Naju, Korea}}
\newcommand{\INSTFH}{\affiliation{Duke University, Department of Physics, Durham, North Carolina, U.S.A.}}
\newcommand{\INSTBA}{\affiliation{Ecole Polytechnique, IN2P3-CNRS, Laboratoire Leprince-Ringuet, Palaiseau, France }}
\newcommand{\INSTEF}{\affiliation{ETH Zurich, Institute for Particle Physics, Zurich, Switzerland}}
\newcommand{\INSTEG}{\affiliation{University of Geneva, Section de Physique, DPNC, Geneva, Switzerland}}
\newcommand{\INSTDG}{\affiliation{H. Niewodniczanski Institute of Nuclear Physics PAN, Cracow, Poland}}
\newcommand{\INSTCB}{\affiliation{High Energy Accelerator Research Organization (KEK), Tsukuba, Ibaraki, Japan}}
\newcommand{\INSTED}{\affiliation{Institut de Fisica d'Altes Energies (IFAE), Bellaterra (Barcelona), Spain}}
\newcommand{\INSTEC}{\affiliation{IFIC (CSIC \& University of Valencia), Valencia, Spain}}
\newcommand{\INSTEI}{\affiliation{Imperial College London, Department of Physics, London, United Kingdom}}
\newcommand{\INSTGF}{\affiliation{INFN Sezione di Bari and Universit\`a e Politecnico di Bari, Dipartimento Interuniversitario di Fisica, Bari, Italy}}
\newcommand{\INSTNA}{\affiliation{INFN Sezione di Napoli and Universit\`a di Napoli, Dipartimento di Fisica, Napoli, Italy}}
\newcommand{\INSTPA}{\affiliation{INFN Sezione di Padova and Universit\`a di Padova, Dipartimento di Fisica, Padova, Italy}}
\newcommand{\INSTBD}{\affiliation{INFN Sezione di Roma and Universit\`a di Roma "La Sapienza", Roma, Italy}}
\newcommand{\INSTEB}{\affiliation{Institute for Nuclear Research of the Russian Academy of Sciences, Moscow, Russia}}
\newcommand{\INSTCC}{\affiliation{Kobe University, Kobe, Japan}}
\newcommand{\INSTCD}{\affiliation{Kyoto University, Department of Physics, Kyoto, Japan}}
\newcommand{\INSTEJ}{\affiliation{Lancaster University, Physics Department, Lancaster, United Kingdom}}
\newcommand{\INSTFC}{\affiliation{University of Liverpool, Department of Physics, Liverpool, United Kingdom}}
\newcommand{\INSTFI}{\affiliation{Louisiana State University, Department of Physics and Astronomy, Baton Rouge, Louisiana, U.S.A.}}
\newcommand{\INSTJ}{\affiliation{Universit\'e de Lyon, Universit\'e Claude Bernard Lyon 1, IPN Lyon (IN2P3), Villeurbanne, France}}
\newcommand{\INSTCE}{\affiliation{Miyagi University of Education, Department of Physics, Sendai, Japan}}
\newcommand{\INSTFJ}{\affiliation{State University of New York at Stony Brook, Department of Physics and Astronomy, Stony Brook, New York, U.S.A.}}
\newcommand{\INSTCF}{\affiliation{Osaka City University, Department of Physics, Osaka,  Japan}}
\newcommand{\INSTGG}{\affiliation{Oxford University, Department of Physics, Oxford, United Kingdom}}
\newcommand{\INSTBB}{\affiliation{UPMC, Universit\'e Paris Diderot, CNRS/IN2P3, Laboratoire de Physique Nucl\'eaire et de Hautes Energies (LPNHE), Paris, France}}
\newcommand{\INSTGC}{\affiliation{University of Pittsburgh, Department of Physics and Astronomy, Pittsburgh, Pennsylvania, U.S.A.}}
\newcommand{\INSTFA}{\affiliation{Queen Mary, University of London, School of Physics and Astronomy, London, United Kingdom}}
\newcommand{\INSTE}{\affiliation{University of Regina, Department of Physics, Regina, Saskatchewan, Canada}}
\newcommand{\INSTGD}{\affiliation{University of Rochester, Department of Physics and Astronomy, Rochester, New York, U.S.A.}}
\newcommand{\INSTBC}{\affiliation{RWTH Aachen University, III. Physikalisches Institut, Aachen, Germany}}
\newcommand{\INSTDD}{\affiliation{Seoul National University, Department of Physics and Astronomy, Seoul, Korea}}
\newcommand{\INSTFB}{\affiliation{University of Sheffield, Department of Physics and Astronomy, Sheffield, United Kingdom}}
\newcommand{\INSTDI}{\affiliation{University of Silesia, Institute of Physics, Katowice, Poland}}
\newcommand{\INSTDA}{\affiliation{STFC, Daresbury Laboratory, Warrington, United Kingdom}}
\newcommand{\INSTEH}{\affiliation{STFC, Rutherford Appleton Laboratory, Harwell Oxford, United Kingdom}}
\newcommand{\INSTCH}{\affiliation{University of Tokyo, Department of Physics, Tokyo, Japan}}
\newcommand{\INSTBJ}{\affiliation{University of Tokyo, Institute for Cosmic Ray Research, Kamioka Observatory, Kamioka, Japan}}
\newcommand{\INSTCG}{\affiliation{University of Tokyo, Institute for Cosmic Ray Research, Research Center for Cosmic Neutrinos, Kashiwa, Japan}}
\newcommand{\INSTF}{\affiliation{University of Toronto, Department of Physics, Toronto, Ontario, Canada}}
\newcommand{\INSTB}{\affiliation{TRIUMF, Vancouver, British Columbia, Canada}}
\newcommand{\INSTG}{\affiliation{University of Victoria, Department of Physics and Astronomy, Victoria, British Columbia, Canada}}
\newcommand{\INSTDJ}{\affiliation{University of Warsaw, Faculty of Physics, Warsaw, Poland}}
\newcommand{\INSTDH}{\affiliation{Warsaw University of Technology, Institute of Radioelectronics, Warsaw, Poland}}
\newcommand{\INSTFD}{\affiliation{University of Warwick, Department of Physics, Coventry, United Kingdom}}
\newcommand{\INSTGE}{\affiliation{University of Washington, Department of Physics, Seattle, Washington, U.S.A.}}
\newcommand{\INSTGH}{\affiliation{University of Winnipeg, Department of Physics, Winnipeg, Manitoba, Canada}}
\newcommand{\INSTEA}{\affiliation{Wroclaw University, Faculty of Physics and Astronomy, Wroclaw, Poland}}
\newcommand{\INSTH}{\affiliation{York University, Department of Physics and Astronomy, Toronto, Ontario, Canada}}

\INSTC     
\INSTDF    
\INSTEE    
\INSTFE    
\INSTD     
\INSTBR
\INSTGA    
\INSTI     
\INSTCI    
\INSTGB    
\INSTFG    
\INSTCJ    
\INSTFH    
\INSTBA    
\INSTEF    
\INSTEG    
\INSTDG    
\INSTCB    
\INSTED    
\INSTEC    
\INSTEI    
\INSTGF    
\INSTNA    
\INSTPA    
\INSTBD    
\INSTEB    
\INSTCC    
\INSTCD    
\INSTEJ    
\INSTFC    
\INSTFI    
\INSTJ     
\INSTCE    
\INSTFJ    
\INSTCF    
\INSTGG    
\INSTBB    
\INSTGC    
\INSTFA    
\INSTE     
\INSTGD    
\INSTBC    
\INSTDD    
\INSTFB    
\INSTDI    
\INSTDA    
\INSTEH    
\INSTCH    
\INSTBJ    
\INSTCG    
\INSTF     
\INSTB     
\INSTG     
\INSTDJ    
\INSTDH    
\INSTFD    
\INSTGE    
\INSTGH    
\INSTEA    
\INSTH     

\author{K.\,Abe}\thanks{also at IPMU, TODIAS, Univ. of Tokyo, Japan}\INSTBJ
\author{N.\,Abgrall}\INSTEG
\author{Y.\,Ajima}\thanks{also at J-PARC Center}\INSTCB
\author{H.\,Aihara}\thanks{also at IPMU, TODIAS, Univ. of Tokyo, Japan}\INSTCH
\author{J.B.\,Albert}\INSTFH
\author{C.\,Andreopoulos}\INSTEH
\author{B.\,Andrieu}\INSTBB
\author{M.D.\,Anerella}\INSTBR
\author{S.\,Aoki}\INSTCC
\author{O.\,Araoka}\thanks{also at J-PARC Center}\INSTCB
% ...

\collaboration{The T2K Collaboration}\noaffiliation

\date{\today}

\begin{abstract}
We report a measurement of muon-neutrino disappearance in the T2K
experiment. The 295-km muon-neutrino beam from Tokai to Kamioka is the
first implementation of the off-axis technique in a
long-baseline neutrino oscillation experiment. 
With data corresponding to 1.43$\times$10$^{20}$ protons on target, we observe 31 
fully-contained single $\mu$-like ring events in Super-Kamiokande, 
compared with an expectation of 104 $\pm$ 14~(syst) 
events without neutrino oscillations. The best-fit point for two-flavor 
$\nu_{\mu} \rightarrow \nu_{\tau}$ oscillations 
is
$\sin^{2}(2 \theta_{23})$ = 0.98 and $|\Delta m_{32}^{2}|$ = 
2.65 $\times$ $10^{-3}$ eV$^{2}$. The boundary of the
90\% confidence region 
includes the points ($\sin^{2}(2 \theta_{23})$, $|\Delta m_{32}^{2}|$) 
= (1.0, 3.1$\times$10$^{-3}$eV$^{2}$), (0.84, 2.65$\times$10$^{-3}$eV$^{2}$)
and (1.0, 2.2$\times$10$^{-3}$eV$^{2}$).
\end{abstract}


\maketitle

We report a measurement of muon-neutrino disappearance in the T2K
experiment. The muon-neutrino beam from Tokai to Kamioka is the
first implementation of the off-axis technique~\cite{beavis:bnl} in a
long-baseline neutrino oscillation experiment. The off-axis technique
is used to provide a narrow-band neutrino energy spectrum 
tuned to the value of $L/E$ that maximizes the neutrino oscillation
effect due to $\Delta m^2_{32}$, the mass splitting first observed in
atmospheric neutrinos~\cite{Fukuda:1998mi}. This narrow-band energy spectrum also
provides a clean signature for subdominant electron neutrino
appearance, as we have recently reported~\cite{Abe:2011sj}. Muon-neutrino 
disappearance depends on the survival probability, which, in the framework of two-flavor
$\nu_{\mu} \rightarrow \nu_{\tau}$ oscillations, is given by
\begin{equation}
P_{surv}= 1-\sin^2 (2 \theta_{23}) \: \sin^2 \left({{\Delta m^{2}_{32} L} \over {4E}} \right), \label{eq:surv}
\end{equation}
where $ E$ is the neutrino energy and $L$ is the neutrino propagation length. 
We have neglected subleading oscillation terms. 
In this paper we describe our observation of 
 $\nu_{\mu}$ disappearance, and we use the result 
to measure $|\Delta m^{2}_{32}|$ and $\sin^2
(2\theta_{23})$. Previous measurements of these neutrino mixing parameters
have been reported by K2K~\cite{Ahn:2006K2K} and MINOS~\cite{Adamson:2011ig}, 
which use on-axis
neutrino beams, and Super-Kamiokande~\cite{sk-2011}, which uses atmospheric
neutrinos.

Details of the T2K experimental setup are described
elsewhere~\cite{Abe:2011ks}. Here we briefly review the components
relevant for the $\nu_\mu$ oscillation analysis.  The J-PARC Main Ring
(MR) accelerator~\cite{cite:Jparc} provides 30~GeV protons with a cycle
of 0.3~Hz. Six bunches (Run 1) or eight bunches (Run 2) are 
extracted in a 5-$\mu$s spill and
are transported to the production target through an 
arc instrumented by superconducting
magnets.  
The proton beam position, profile, timing and intensity are measured by
21 electrostatic beam position monitors
(ESM), 19 segmented secondary emission monitors (SSEM), one optical
transition radiation monitor (OTR) and five current transformers.  The
secondary beamline, filled with helium at atmospheric pressure, is
composed of the target, focusing horns and decay tunnel. The graphite
target is 2.6~cm in diameter and 90~cm (1.9 $\lambda_{int}$) long.
Positively-charged particles exiting the target are focused
into the 96-m long decay tunnel by three magnetic horns pulsed at
250~kA.  Neutrinos are primarily produced in the decays of charged
pions and kaons.  A beam dump is located at the end of the tunnel and
is followed by muon monitors measuring the beam direction of each spill.

The neutrino beam is directed
2.5$^\circ$ off the axis between the target and 
the Super-Kamiokande (SK) far detector 295 km away. 
This configuration produces a narrow-band \numu
beam with peak energy tuned to the first oscillation maximum
$E_{\nu}=|\Delta m^{2}_{32}| L/(2\pi)\simeq$ 0.6~GeV. 

The near detector complex (ND280)~\cite{Abe:2011ks} is located 280~m downstream from the target and hosts
two detectors.
The on-axis Interactive Neutrino GRID (INGRID)~\cite{ingrid-nim} records neutrino interactions with high statistics to monitor the beam intensity,
direction and profile. It consists of 14 identical 7-ton modules
composed of an iron-absorber/scintillator-tracker sandwich arranged in 10~m by 10~m crossed horizontal and vertical arrays centered on the beam.
The off-axis detector 
reconstructs exclusive final states to study neutrino interactions
and beam properties corresponding to those expected at the far detector. 
Embedded  in the refurbished UA1/NOMAD magnet (field strength 0.2~T), it consists of three 
large-volume time projection chambers (TPCs)~\cite{Abgrall:2010hi} interleaved
with two fine-grained tracking detectors (FGDs, each 1~ton).
It also has a $\pi^0$-optimized detector and a surrounding electromagnetic calorimeter.
The magnet yoke is instrumented as 
a side muon range detector.

The SK water-Cherenkov far detector~\cite{fukuda:2002uc}
has a fiducial volume (FV) of  22.5~kt within its cylindrical inner detector (ID).
Enclosing the ID is the 2 m-wide outer detector (OD).
The front-end readout electronics~\cite{Abe:2011ks} allow for a 
dead-time-free trigger.
Spill timing information, synchronized by the Global Positioning System (GPS) 
with $<150$~ns precision,
is transferred from J-PARC
to SK and triggers the recording of
photomultiplier (PMT) hits within $\pm$500 $\mu$s of the expected neutrino arrival time.

The results presented in this Letter are based on the first two 
physics runs: Run~1 (Jan--Jun 2010) and Run~2 (Nov 2010--Mar 2011).
During this time period, the MR proton beam power was continually increased and 
reached
145~kW with $9\times 10^{13}$ protons per pulse.
The 
fraction of protons hitting the target was monitored by the ESM, SSEM and OTR and found to be greater than 99\% and stable in time.
A total of 2,474,419 spills was retained for analysis after beam and far-detector 
quality cuts, corresponding to 
$1.43\times10^{20}$ protons on target (POT).

We present the study of events in the far detector with a single muon-like ($\mu$-like) ring.
The event selection enhances $\nu_{\mu}$ charged-current quasi-elastic interactions (CCQE).  For these events, neglecting the Fermi motion, the neutrino energy $E_{\nu} $ can be reconstructed as 
 \begin{equation}\label{eqn:erecccqe}
E_{\nu}= {{m^2_p-(m_{n}-E_b)^2-m^2_{\mu}+ 2 (m_n -E_b) E_{\mu} } \over {2(m_{n}-E_b-E_{\mu}+p_{\mu}\cos \theta_{\mu})}},
\end{equation}
where $m_p$ is the proton mass, $m_{n}$ the neutron mass, and $E_b=27$~MeV
the binding energy of a nucleon inside a $^{16}$O nucleus. In Eq.~\ref{eqn:erecccqe} $E_{\mu}$, $p_{\mu}$, 
and $\theta_{\mu}$ are respectively the measured muon energy, momentum and angle with respect to the incoming neutrino.
The selection criteria for this analysis were fixed from Monte Carlo (MC) studies before the data 
were collected.
The observed number of events and spectrum are compared with signal and background expectations, which are based on 
neutrino flux and cross-section predictions and are
corrected using an inclusive measurement in the 
off-axis near detector.

\begin{figure}[!tbp]
  \centering
  \includegraphics[width=2.9in]{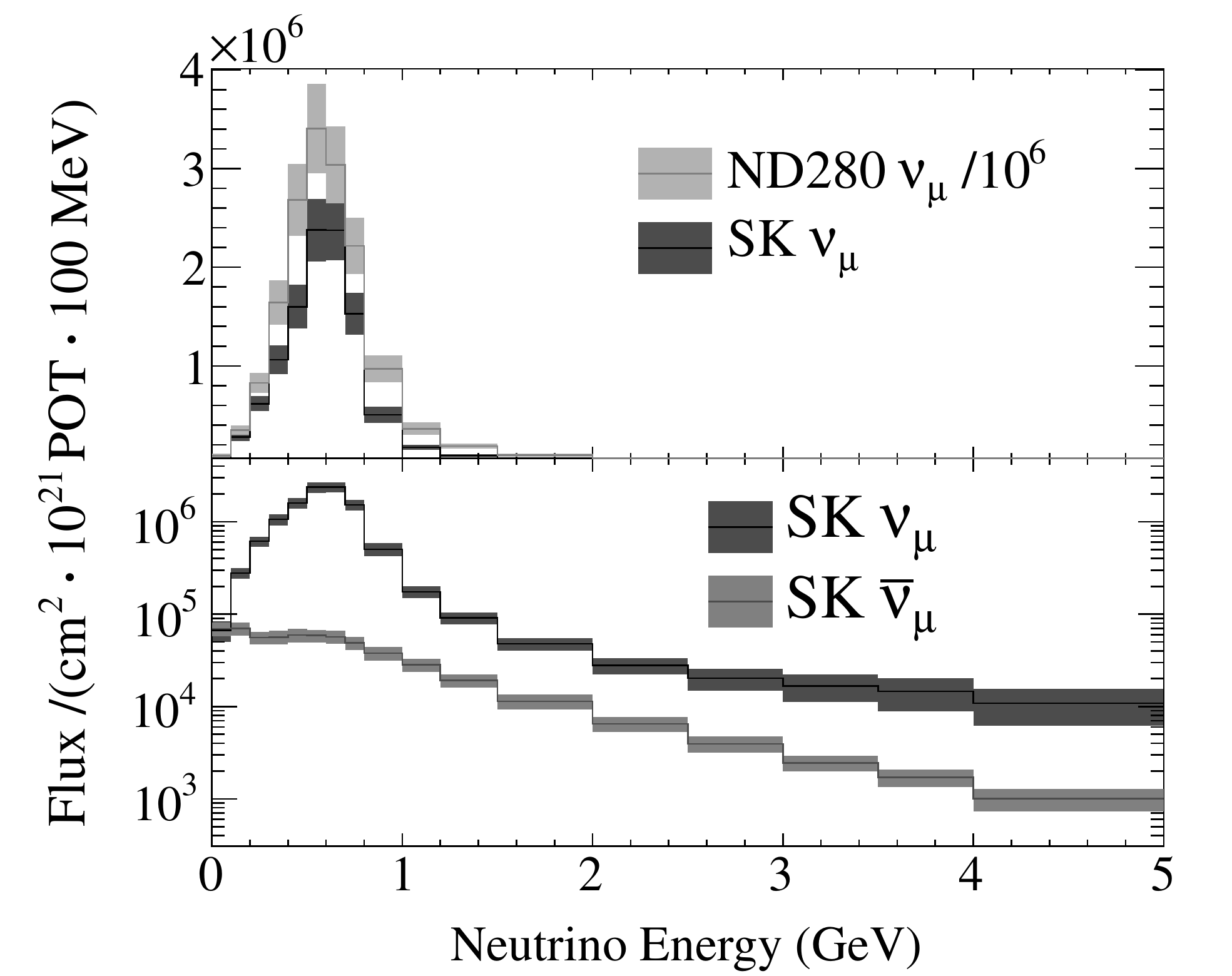}
 \vspace{-0.5cm}
  \caption{(Top) The predicted flux of $\nu_\mu$  as a function of neutrino energy without oscillations at Super-Kamiokande and at the off-axis near detector; (Bottom) the flux of $\nu_\mu$ and $\overline{\nu}_\mu$ at Super-Kamiokande. The shaded boxes indicate the total systematic uncertainty for each energy bin.}
  \label{fig:beamflux}
\end{figure}

Our predicted beam flux (Fig.~\ref{fig:beamflux}) is based on models 
tuned to experimental data.  The most
significant constraint comes from NA61 measurements of pion
production~\cite{Abgrall:2011ae} 
in ($p$, $\theta$) bins, where $p$ is the pion momentum and $\theta$ the polar angle with respect to the proton beam; 
there are 5\%-10\% systematic and similar statistical uncertainties in most
of the measured phase space. The production of pions in the target
outside the NA61-measured phase space
and all kaon production are modeled using FLUKA \cite{fluka1,fluka2}.
The production rate of these pions is assigned systematic uncertainties
of 50\%, and kaon production uncertanties are estimated to be between
15\% and 100\% based on a comparison of FLUKA with data from
Eichten et al.~\cite{Eichten}. 
The software package GEANT3~\cite{GEANT3},
with GCALOR~\cite{GCALOR}
for hadronic interactions, handles particle propagation through the
magnetic horns, target hall, decay volume and beam dump.  Additional
systematic errors in the neutrino fluxes are included for uncertainties
in secondary nucleon production and total hadronic inelastic
cross sections, uncertainties in the proton beam direction, spatial
extent and angular divergence, the horn current, 
and the secondary beam line component
alignment uncertainties.  The stability of the beam direction and
neutrino rate per proton on target are monitored continuously with
INGRID and are within the assigned systematic
uncertainties~\cite{Abe:2011sj}.

Systematic uncertainties in the shape of the flux as
a function of neutrino energy require knowledge of the correlations of
the uncertainties in ($p$, $\theta$) bins of hadron production.  
For the NA61 pion-production
data~\cite{Abgrall:2011ae}, we assume full correlation
between ($p$, $\theta$) bins for each individual source of systematic
uncertainty, except for particle identification where there is a known
momentum-dependent correlation.  Where correlations of 
hadron-production uncertainties are unknown, we choose correlations in
kinematic variables to maximize the uncertainty in the normalization of 
the predicted flux.

Neutrino interactions are simulated using the NEUT event generator
\cite{hayato:neut2}. 
Uncertainties in 
cross sections of the exclusive neutrino processes 
are determined by comparisons with recent measurements from the 
SciBooNE \cite{sciboone:ccqe}, MiniBooNE 
\cite{miniboone:ccqe, miniboone:cc1pirat}, 
and K2K \cite{Gran:2006K2K, Rodriguez:2008K2K} 
experiments, comparisons with the GENIE \cite{Andreopoulos:2009rq} 
and NuWro \cite{Juszczak:2009} generators 
and recent theoretical work \cite{Juszczak:2010}.
 
\begin{figure}[!htbp]
  \centering
  \includegraphics[width=2.9in]{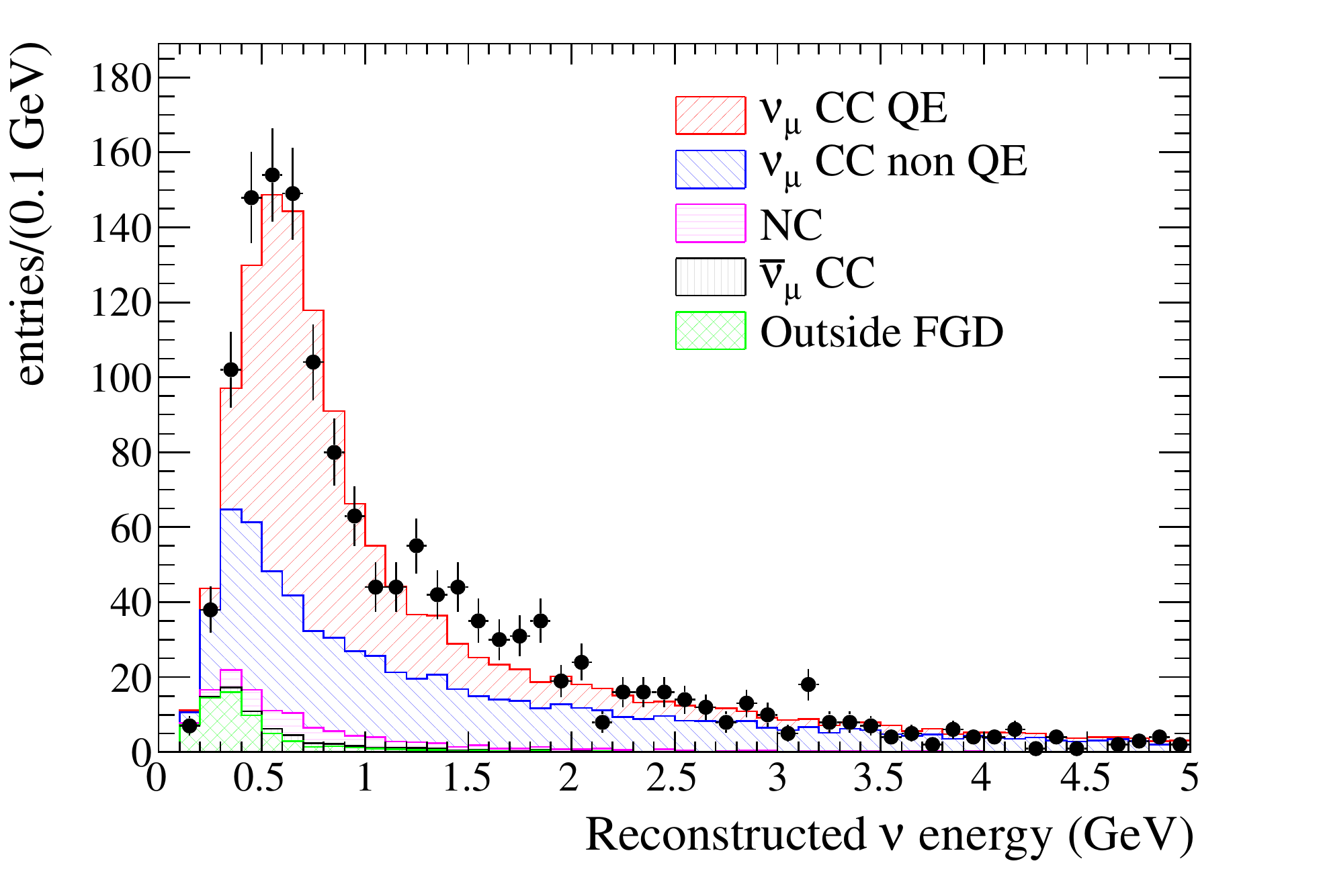}
  \caption{
Neutrino energy reconstructed for the CCQE hypothesis for $\nu_\mu$~CC candidates interacting in the FGD target.
  The data are shown using points with error bars (statistical only) and the MC predictions are in 
  shaded histograms.
   }
  \label{fig:ND280momentum}
\end{figure}

An inclusive $\nu_\mu$ charged-current (CC) measurement in the off-axis near detector (ND)
is used to constrain the expected event rate at the far detector.
From a data sample collected in Run~1 of $2.88\times10^{19}$ POT,
neutrino interactions are selected in the FGDs with charged particles entering the downstream TPC.
The most energetic negatively charged particle in the TPC is required 
to have ionization energy loss compatible with that of a muon.
The analysis selects 1529 data events with 38\% $\nu_\mu$~CC efficiency and 90\% purity. The agreement between the reconstructed neutrino energy in data and MC is shown in Fig.~\ref{fig:ND280momentum}.
The ratio 
of measured $\nu_\mu$ CC interactions to MC is
\begin{eqnarray}
R^{\nu_{\mu} CC}_{ND} &=& \frac{N^{Data,\nu_{\mu} CC}_{ND}}{N^{MC, \nu_{\mu} CC}_{ND}} = 1.036 \pm 0.028 (\mathrm{stat.}) \nonumber \\
&& ^{+0.044}_{-0.037} (\mathrm{det.syst.}) \pm 0.038(\mathrm{phys.syst.}),
\label{eq:ratiodmc}
\end{eqnarray}
where $N^{Data,\nu_{\mu} CC}_{ND}$ is the number of $\nu_{\mu}$ CC events, and $N^{MC, \nu_{\mu} CC}_{ND}$ is the MC prediction normalized by POT. 
The detector systematic errors in Eq.~\ref{eq:ratiodmc} are mainly due to uncertainties in tracking and particle identification
efficiencies. 
The physics uncertainties result from cross section uncertainties 
but exclude normalization uncertainties that cancel in a far/near ratio.

At the far detector we select a $\nu_\mu$ CCQE enriched sample.
The SK event reconstruction~\cite{ashie:2005ik} 
uses PMT hits 
in time with a neutrino spill.
We select a fully-contained fiducial volume (FCFV) sample by requiring 
no activity in the OD,  
no pre-activity in the 100 $\mu$s before the event trigger time,
at least $30$ MeV electron-equivalent energy deposited in the ID,
and a reconstructed event vertex in the fiducial region.
The OD veto rejects events induced by neutrino interactions outside of the ID,
and events where energy escapes from the ID. The visible energy requirement rejects events from radioactive decays in the detector.  The fiducial vertex requirement rejects particles entering from outside the ID. 
Further conditions are required to enrich the sample in $\nu_\mu$ CCQE events:
a single Cherenkov ring identified as a muon, with momentum $p_\mu>200$ MeV/c, and no more than one delayed electron.
The muon momentum requirement rejects charged pions and misidentified electrons from the decay of unseen muons and pions, and the delayed-electron veto rejects events with muons accompanied by unseen pions and muons.
The number of events in data and MC after each selection criterion is shown in Table~\ref{table:number_of_events}.
The efficiency and purity of $\nu_{\mu}$ CCQE events are estimated to be 72\% and 61\% respectively.

 \begin{table}[!tbp]
  \centering
  \caption{Event reduction 
at the far detector.
   After each selection criterion is applied, the 
 number of observed (Data) and MC expected events of
 $\nu_\mu$~CCQE,  $\nu_\mu$~CC non-QE, intrinsic \nue, and neutral current (NC) are given.
 The columns denoted by \numu include \numubar.
 All MC CC samples assume $\nu_\mu \rightarrow \nu_\tau$ oscillations with
 $\sin^2 (2\theta_{23})$=1.0 and $|\Delta{m}^2_{32}|$=$2.4\times10^{-3}$eV$^2$.}
  \begin{tabular}{lccccc}
   \hline
   \hline
                        & Data & \numu{}CCQE & \numu{}CC non-QE & \nue{}CC & NC  \\
   \hline
   FV interaction                &  n/a    & 24.0 & 43.7 &  3.1      & 71.0 \\
   FCFV              & 88      & 19.0 & 33.8 &  3.0      & 18.3 \\
   single ring                     & 41      & 17.9 & 13.1 &  1.9      & 5.7 \\
   $\mu$-like                      & 33      & 17.6 & 12.4 &  $<$0.1   & 1.9 \\
   $p_{\mu}>200$~MeV/c               & 33      & 17.5 & 12.4 &  $<$0.1   & 1.9 \\
   0 or 1 delayed $e$    & 31      & 17.3 &  9.2 &  $<$0.1   & 1.8 \\
   \hline
   \hline
  \end{tabular}
  \label{table:number_of_events}
 \end{table}

We calculate the expected number of signal events in the far detector
($N_{SK}^{exp}$) by correcting the far-detector MC prediction with $R^{\nu_{\mu} CC}_{ND}$ from
Eq. \ref{eq:ratiodmc}:
\begin{equation}
N_{SK}^{exp}(E_{r}) = R^{\nu_{\mu} CC}_{ND}
\sum_{E_{t}} P_{surv}(E_{t}) N_{SK}^{MC}(E_{r},E_{t}).
\label{eq:nskexp}
\end{equation}
In Eq.~\ref{eq:nskexp}, $N_{SK}^{MC}(E_{r},E_{t})$ is the expected number of events for the no-disappearance hypothesis for T2K 
Runs 1 and 2 in bins of reconstructed ($E_{r}$) and true ($E_{t}$) energies. $P_{surv}(E_{t})$
is the two-flavor $\nu_{\mu}$-survival probability, and is applied to $\nu_{\mu}$ and $\bar{\nu}_{\mu}$ CC interactions but not 
to neutral-current interactions.

The sources of systematic uncertainty in $N_{SK}^{exp}$ are listed in Table~\ref{table:nsksystematics}. 
Uncertainties in the near-detector and far-detector selection efficiencies
are energy-independent except for 
the ring-counting efficiency.
Uncertainty in the near-detector event rate is
applied to $N^{Data,\nu_{\mu} CC}_{ND}$ in Eq. \ref{eq:ratiodmc}. 
The flux normalization uncertainty is reduced because of the near-detector constraint.
The uncertainty in the flux shape is propagated using the covariance matrix when calculating $N_{SK}^{exp}$.
The near-detector constraint also leads to partial cancellation in the uncertainty 
in cross section modeling, but the cancellation is not complete due to the different fluxes, 
different acceptances and different nuclei in the near and far detectors.
The total uncertainty in $N_{SK}^{exp}$ is $^{+13.3\%}_{-13.0\%}$ without oscillations and $^{+15.0\%}_{-14.8\%}$ 
with oscillations with $\sin^{2}(2 \theta_{23})$ = 1.0 and $|\Delta m_{32}^{2}|$ = 2.4 $\times$ $10^{-3}$ eV$^{2}$. 
 \begin{table}[!tbp]
  \centering
  \caption{Systematic uncertainties on the predicted number of SK selected events without oscillations and for oscillations with 
   $\sin^{2}(2 \theta_{23})$ = 1.0 and $|\Delta m_{32}^{2}|$ = 2.4 $\times$ $10^{-3}$ eV$^{2}$.}
  \begin{tabular}{lcc}
   \hline
   \hline
   Source                     &$\delta N_{SK}^{exp}/N_{SK}^{exp}$ & $\delta N_{SK}^{exp}/N_{SK}^{exp}$ \\
                              & (\%, no osc) & (\%, with osc) \\
   \hline
   SK CCQE efficiency               & $\pm 3.4$    &  $\pm 3.4$ \\
   SK CC non-QE efficiency          &  $\pm 3.3$   & $\pm 6.5$   \\
   SK NC efficiency                 &  $\pm 2.0$   & $\pm 7.2$   \\
   ND280 efficiency                 &  +5.5 -5.3   & +5.5 -5.3   \\
   ND280 event rate                 &  $\pm 2.6$   & $\pm 2.6$ \\
   Flux normalization (SK/ND280)    &  $\pm 7.3$   & $\pm 4.8$ \\
   CCQE cross section               &  $\pm 4.1$   & $\pm 2.5$ \\
   CC1$\pi$/CCQE cross section      &  +2.2 -1.9   & +0.4 -0.5  \\
   Other CC/CCQE cross section      &  +5.3 -4.7   & +4.1 -3.6 \\
   NC/CCQE cross section            &  $\pm 0.8$   & $\pm 0.9$ \\
   Final-state interactions         &  $\pm 3.2$   & $\pm 5.9$ \\
   \hline
   Total                            & +13.3 -13.0  & +15.0 -14.8  \\
   \hline
   \hline
  \end{tabular}
  \label{table:nsksystematics}
 \end{table}

We find the best-fit values of the oscillation parameters using a binned likelihood-ratio method, in which $\sin^{2}(2 \theta_{23})$ 
and $|\Delta m_{32}^{2}|$ are varied in the input to the calculation of $N_{SK}^{exp}$ until 
\begin{equation}
2  \sum_{E_{r}} \left [ N_{SK}^{data}  \ln \left ( \frac{N_{SK}^{data}}{N_{SK}^{exp}} \right ) + (N_{SK}^{exp} - N_{SK}^{data}) \right ]
\label{eq:chisq}
\end{equation}
is minimized. The sum in Eq.\ \ref{eq:chisq} is over 50 MeV bins of reconstructed energy of selected events in the far detector from 
0-10 GeV.

\begin{figure}[!tbp]
  \centering
  \includegraphics[width=2.9in]{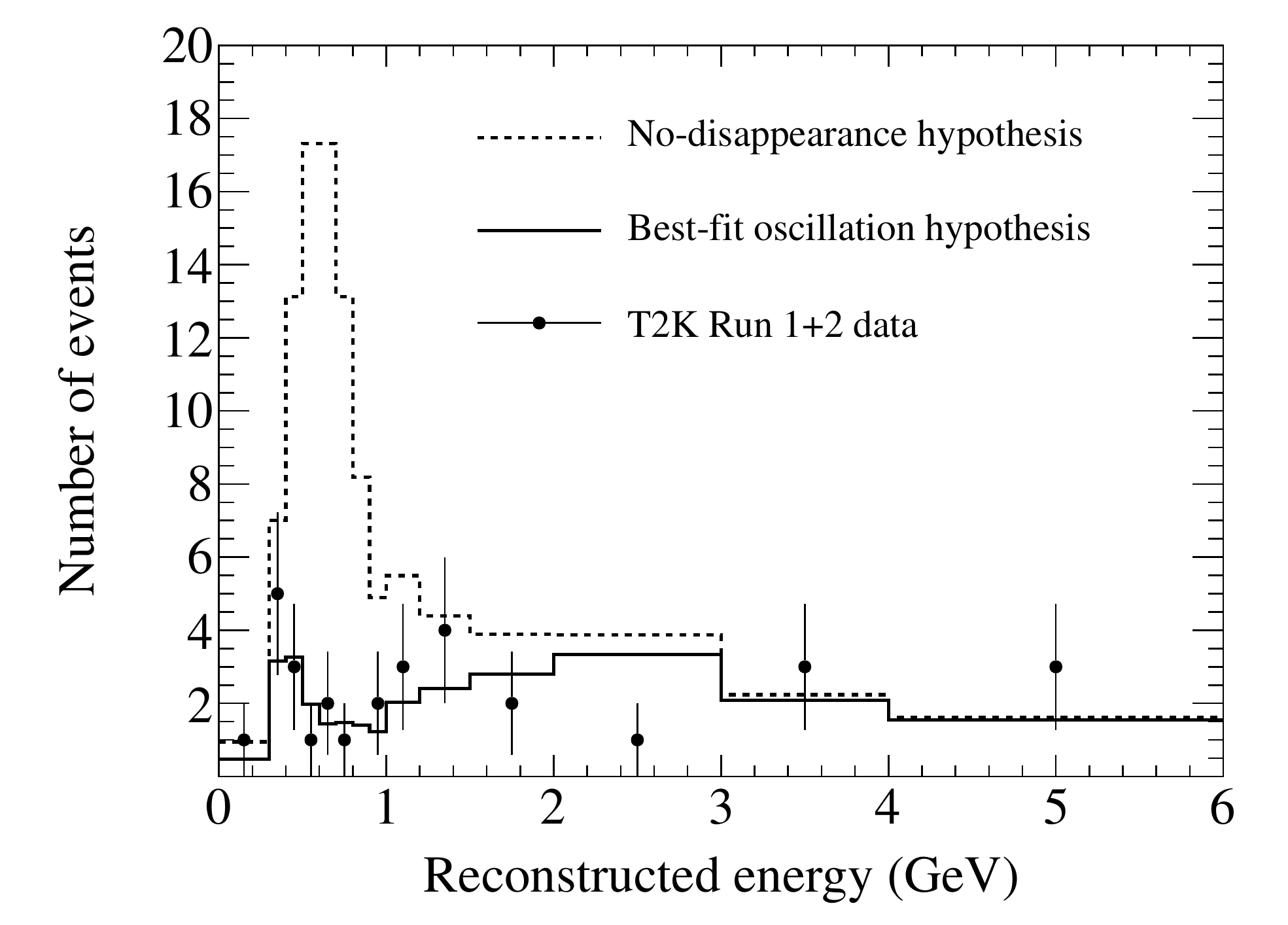}
 \vspace{-0.5cm}
  \caption{Reconstructed energy spectrum of the 31 data events compared with the expected spectra in the far detector without disappearance and 
with best-fit $\nu_{\mu} \rightarrow \nu_{\tau}$ oscillations.
A variable binning scheme is used here for
 the purpose of illustration only; 
the actual analysis used equal-sized 50 MeV bins.}
  \label{fig:recospectra}
\end{figure}

Using the near-detector measurement and setting $P_{surv}$ = 1.0 in Eq.\ \ref{eq:nskexp}, we 
expect a total of 
103.6 
$^{+13.8}_{-13.4}$ (syst) single $\mu$-like ring events in the far 
detector without disappearance, but we observe 31 events. 
If 
$\nu_{\mu} \rightarrow \nu_{\tau}$ oscillations are assumed, the 
best-fit point determined using Eq.\ \ref{eq:chisq} is $\sin^{2}(2 \theta_{23})$ = 0.98 and 
$|\Delta m_{32}^{2}|$ = 2.65 $\times$ $10^{-3}$ eV$^{2}$. We estimate the systematic uncertainty in the best-fit value 
of $\sin^{2}(2 \theta_{23})$ to be $\pm$4.7\% and that in $|\Delta m_{32}^{2}|$ to be $\pm$4.5\%.
The reconstructed energy spectrum of the 31 data events is shown in Fig.\ \ref{fig:recospectra} along with the expected 
far-detector spectra without disappearance and with best-fit oscillations.

\begin{figure}[!tbp]
  \centering
  \includegraphics[width=2.9in]{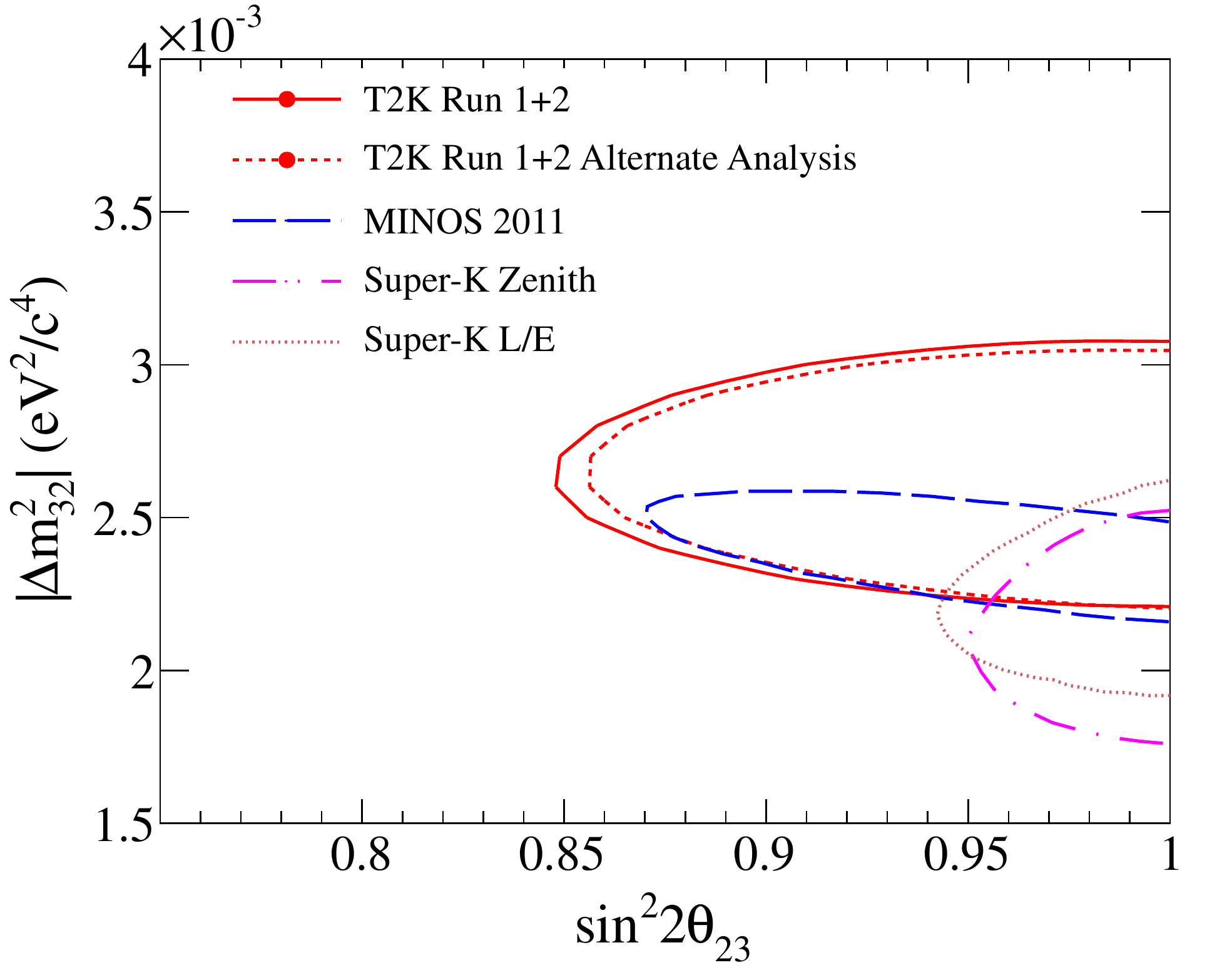}
 \vspace{-0.5cm}
  \caption{The 90\% confidence regions for $\sin^{2}(2 \theta_{23})$ and $|\Delta m_{32}^{2}|$; results
          from the two analyses reported here are compared with those from MINOS \cite{Adamson:2011ig} and Super-Kamiokande 
          \cite{sk-2011, takeuchi}.}
  \label{fig:fccontours}
\end{figure}

We construct confidence regions 
\footnote{
In the T2K narrow-band beam, for a low-statistics data set, there is a
possible degeneracy between the first oscillation maximum and
other oscillation maxima in $L/E$. Therefore we decided in advance to report
confidence regions both with and without an explicit bound at
$|\Delta m_{32}^2|<5\times 10^{-3}$eV$^2$.  For this data set, 
the bounded and unbounded confidence regions are
identical.
}
in the oscillation parameters using the method of Feldman and 
Cousins \cite{cite:feldman_cousins}. Statistical variations are taken into account by Poisson fluctuations
of toy MC datasets, and systematic uncertainties are incorporated using the method of Cousins and 
Highland \cite{cite:Cousins_Highland, cite:conradetal}. The 90\% confidence region for $\sin^{2}(2 \theta_{23})$ 
and $|\Delta m_{32}^{2}|$ is shown in Fig.\ \ref{fig:fccontours} for combined statistical and systematic uncertainties.

We also carried out an alternate analysis with a maximum likelihood method. The likelihood is defined as: 
\begin{eqnarray}\label{eqn:LikelihoodAnalysisA}
L&=&L_{\mbox{norm}}(\sin^2(2\theta_{23}),\Delta{m_{32}^2},{\bf f}) \nonumber \\
&& L_{\mbox{shape}}(\sin^2(2\theta_{23}),\Delta{m_{32}^2},{ \bf f})  
L_{\mbox{syst}}({\bf f}),
\end{eqnarray}
where the first term is the Poisson probability for the observed number of events, and
the second term is the unbinned likelihood for the reconstructed neutrino energy spectrum.
The vector ${ \bf f}$ represents parameters related to systematic uncertainties that have been allowed to vary in the fit to maximize the likelihood, 
and the last term in Eq.~\ref{eqn:LikelihoodAnalysisA} is
a multidimensional Gaussian probability for the systematic error parameters.
The result is consistent with the analysis described earlier.
The 
best-fit point for this alternate analysis is $\sin^{2}(2 \theta_{23})$
 = 0.99 and 
$|\Delta m_{32}^{2}|$ = 2.63 $\times$ $10^{-3}$ eV$^{2}$. The 90\% confidence region for the neutrino oscillation parameters 
is shown in Fig.~\ref{fig:fccontours}.

In conclusion, we have reported the first observation of $\nu_{\mu}$ disappearance using detectors positioned 
off-axis in the beam of a long-baseline neutrino experiment. The values of the oscillation parameters 
$\sin^{2}(2 \theta_{23})$ and $|\Delta m_{32}^{2}|$ obtained are consistent with those reported by 
MINOS \cite{Adamson:2011ig} and Super-Kamiokande \cite{sk-2011, takeuchi}.

\vspace{0.5cm}
\begin{acknowledgments}
We thank the J-PARC accelerator team for the superb accelerator performance and
CERN NA61 colleagues for providing essential particle production data and for their fruitful collaboration.
We acknowledge the support of MEXT, Japan; 
NSERC, NRC and CFI, Canada;
CEA and CNRS/IN2P3, France;
DFG, Germany; 
INFN, Italy;
Ministry of Science and Higher Education, Poland;
RAS, RFBR and the Ministry of Education and Science
of the Russian Federation; 
MEST and NRF, South Korea;
MICINN and CPAN, Spain;
SNSF and SER, Switzerland;
STFC, U.K.; NSF and 
DOE, U.S.A.
We also thank CERN for their donation of the UA1/NOMAD magnet 
and DESY for the HERA-B magnet mover system.
In addition, participation of individual researchers
and institutions in T2K has been further supported by funds from: ERC (FP7), EU; JSPS, Japan; Royal Society, UK; 
DOE Early Career program, and the A. P. Sloan Foundation, U.S.A.

\end{acknowledgments}

\bibliographystyle{apsrev4-1} 
\bibliography{references}

\end{document}